\documentclass[11pt,a4paper]{article}

\usepackage{footnote}

\usepackage[T1]{fontenc}
\usepackage[bitstream-charter]{mathdesign}

\usepackage[latin1]{inputenc}
\usepackage{amsmath}
\usepackage{cite}

\usepackage{xcolor}
\definecolor{bl}{rgb}{0.0,0.2,0.6}

\usepackage{graphicx}
\usepackage{sectsty}
\usepackage[compact]{titlesec}
\allsectionsfont{\color{bl}\scshape\selectfont}

\makeatletter
\def\printtitle{%
    {\color{bl} \centering \huge \sc \textbf{\@title}\par}}
\makeatother

\title{\\ \large \vspace*{-10pt} Pseudospin and Spin Symmetric Solutions of Dirac Equation:
Hellmann Potential, Wei Hua Potential, Varshni
Potential\vspace*{10pt}}

\makeatletter
\def\printauthor{%
    {\centering \small \@author}}
\makeatother

\author{%
    Altuð Arda \\
    arda@hacettepe.edu.tr \\
    Ramazan Sever \\
    sever@metu.edu.tr \\
    \vspace{20pt}
    }

\usepackage{fancyhdr}
\usepackage{lastpage}
\usepackage[runin]{abstract}
\abslabeldelim{\quad}
\catcode`ð=\active
 \defð{\u{g}}
 \catcode`Ð=\active
 \defÐ{\u{G}}
 \catcode`Ý=\active
 \defÝ{\. I}
 \catcode`ö=\active
 \defö{\"{o}}
 \catcode`Ö=\active
 \defÖ{\"O}
 \catcode`ü=\active
 \defü{\"{u}}
 \catcode`Ü=\active
 \defÜ{\"{U}}
 \catcode`Þ=\active
 \defÞ{\c{S}}
 \catcode`þ=\active
 \defþ{\c{s}}
 \catcode`ý=\active
 \defý{{\i}}
 \catcode`ç=\active
 \defç{\d{c}}
 \catcode`Ç=\active
 \defÇ{\d{C}}

\makesavenoteenv{tabular}
\begin{document}

\printtitle

\printauthor

\begin{abstract}
Approximate analytical solutions of the Dirac equation are
obtained for the Hellmann potential, Wei Hua potential and Varshni
potential with any $\kappa$-value for the cases having the Dirac
equation pseudospin and spin symmetries. Closed forms of the
energy eigenvalue equations and the spinor wave functions are
obtained by using the Nikiforov-Uvarov method and some tables are
given to see the dependence of the energy eigenvalues on different
quantum number pairs ($n, \kappa$).
\end{abstract}

\section{Introduction}
The pseudospin and spin symmetric solutions of the Dirac equation
have been great interest in literature for last decades
\cite{ref1,ref2,ref3}. The Dirac equation with vector, $V(r)$, and
scalar, $S(r)$, potentials has pseudospin (spin) symmetry when the
difference (the sum) of the potentials $V(r)-S(r)$ $[V(r)+S(r)]$
is constant, which means $\frac{d}{dr}[V(r)-S(r)]=0$ (or
$\frac{d}{dr}[V(r)+S(r)]=0$). It is pointed out that these
symmetries can explain degeneracies in single-particle energy
levels in nuclei or in some heavy meson-spectra within the
contexts of relativistic mean-field theories
\cite{ref1,ref2,ref3}. In the relativistic domain, these
symmetries were used in the context of deformation and
superdeformation in nuclei, magnetic moment interpretation and
identical bands \cite{ref4}. In the non-relativistic domain,
performing a helicity unitary transformation to a single-particle
Hamiltonian maps the normal state onto the pseudo-state
\cite{ref5}. Moreover, the Dirac Hamiltonian has not only a spin
symmetry but also a $U(3)$ symmetry for the case $V(r)=S(r)$ while
not only a pseudospin symmetry but also a pseudo-$U(3)$ symmetry
with vector and scalar harmonic oscillator potentials
\cite{ref6,ref7}. Because of these investigations, the solutions
of the Dirac equation having spin and pseudospin symmetry have
received great attention for different type of potentials such as
Morse, Eckart, the modified P\"{o}schl-Teller, the Manning-Rosen
potentials and the symmetrical well potential
\cite{ref8,ref9,ref10,ref11,ref12,ref13,ref14,ref15}.

Throughout the paper we use the following approximation instead of
the spin-orbit coupling term to obtain the analytical solutions of
the Hellmann potential, Wei Hua and Varshni potentials
\cite{ref16,ref17,ref18,ref19,ref20,ref21,ref22,ref23}
\begin{eqnarray}
\frac{1}{r^2} \approx \beta^2\,\frac{1}{(1-e^{-\beta r})^2}\,,
\end{eqnarray}
where $\beta$ is a parameter related with the above potentials.

The potentials studied in the present work and also other some
exponential-type potentials such as a ring-shaped H\`{u}lthen, the
Yukawa and Tietz-Hua potentials have been analyzed in details by
using different methods
\cite{ref24,ref25,ref26,ref27,ref28,ref29,ref30}. We intend to use
the Nikiforov-Uvarov method (NU) to analyze the bound states of
the Dirac equation for the cases of pseudospin and spin
symmetries. This method is a powerful tool to solve a second-order
differential equation and has been used to find the bound states
of different potentials in literature \cite{ref31,ref32}.

The organization of this work is as follows. In Section 2, we
briefly give the Dirac equation with attractive scalar and
repulsive vector potentials for the cases where the Dirac equation
has pseudospin and spin symmetries, respectively. In Section 3, we
present the NU method and the parameters required within the
method. In Section 4, we find an analytical energy eigenvalue
equations for the bound states and the two-component spinor wave
functions of the above potentials by using an approximation
instead of the spin-orbit coupling term. In Section 5, we give our
results and discussions. The last section includes our
conclusions.

\section{Dirac Equation}

The free particle Dirac equation is given by ($\hbar=c=1$)
\begin{eqnarray}
\bigl(i\gamma^{\mu}\partial_{\mu}-M\bigr)\Psi(\vec{r},t)=0\,,
\end{eqnarray}
Taking the total wave function as
$\Psi(\vec{r},t)=e^{-iEt}\psi(\vec{r})$ for time-independent
potentials, where $E$ is the relativistic energy, $M$ is particle
mass, the Dirac equation with spherical symmetric vector and
scalar potentials is written as
\begin{eqnarray}
\bigl[\vec{\alpha}.\vec{P}+\beta(M+
S(r))\bigr]\psi(\vec{r})=\big[E-V(r)\bigr]\psi(\vec{r})\,,
\end{eqnarray}
Here $\alpha$ and $\beta$ are usual $4\times4$ matrices. For
spherical nuclei, the angular momentum $\vec{J}$ and the operator
$\hat{K}=-\beta\bigl(\hat{\sigma}.\hat{L}+1\bigr)$ with
eigenvalues $\kappa=\pm (j+1/2)$ commute with the Dirac
Hamiltonian, where $\hat{L}$ is the orbital angular momentum. By
using the radial eigenfunctions for upper and lower components of
the Dirac eigenfunction $F(r)$ and $G(r)$, respectively, the wave
function is written as \cite{ref31}
\begin{eqnarray}
\psi(\vec{r})=\,\frac{1}{r}\,\Bigg[\begin{array}{c}
\,F\,(r)Y^{(1)}(\theta,\phi) \\
iG\,(r)Y^{(2)}(\theta,\phi)
\end{array}\Bigg]\,,
\end{eqnarray}
where $Y^{(1)}(\theta,\phi)$ and $Y^{(2)}(\theta,\phi)$ are the
pseudospin and spin spherical harmonics, respectively. They
correspond to angular and spin parts of the wave function given by
\begin{eqnarray}
Y^{(1),(2)}(\theta,\phi)=\sum_{m_{\ell}m_{s}}<\ell
m_{\ell}\frac{1}{2}m_{s}|\ell\frac{1}{2}jm>Y_{\ell
m_{\ell}}(\theta,\phi)\chi_{\frac{1}{2}m_{s}}\,,\nonumber\\j=|\kappa|-\frac{1}{2}\,,\,\,\,
\ell=\kappa\,\,(\kappa>0)\,;\,\ell=-(\kappa+1)\,\,(\kappa<0)\,,
\end{eqnarray}
Here, $Y_{\ell m_{\ell}}(\theta,\phi)$ denotes the spherical
harmonics and $m_{\ell}$ and $m_{s}$ are related magnetic quantum
numbers.

Substituting Eq. (4) into Eq. (3) gives us the following coupled
differential equations
\begin{subequations}
\begin{align}
&&\left(\frac{d}{dr}+\frac{\kappa}{r}\right)F(r)=[E+M-\Gamma(r)]G(r)\,,\\
&&\left(\frac{d}{dr}-\frac{\kappa}{r}\right)G(r)=[M-E+\Lambda(r)]F(r)\,.
\end{align}
\end{subequations}
where $\Gamma(r)=V(r)-S(r)$ and $\Lambda(r)=V(r)+S(r)$. Using the
expression $G(r)$ in Eq. (6a) and inserting it into Eq. (6b), we
get a second order differential equation
\begin{eqnarray}
\left[\frac{d^2}{dr^2}-\frac{\kappa(\kappa+1)}{r^2}+\varepsilon^{(1)}(r)\right]F(r)
=-\left[\frac{d\Gamma(r)/dr}{\left[E+M-\Gamma(r)\right]}\right]F(r)\,,
\end{eqnarray}
where $\varepsilon^{(1)}(r)=\left[E+M-\Gamma(r)\right]
\left[E-M-\Lambda(r)\right]$. By similar steps, we write the
following second order differential equation for $G(r)$ as
\begin{eqnarray}
\left[\frac{d^2}{dr^2}-\frac{\kappa(\kappa-1)}{r^2}+\varepsilon^{(2)}(r)\right]
G(r)=\left[\frac{d\Lambda(r)/dr}{\left[M-E+\Lambda(r)\right]}\right]G(r)\,,
\end{eqnarray}
where $\varepsilon^{(2)}(r)=\left[E-M-\Lambda(r)\right]
\left[E+M-\Gamma(r)\right]$. If the Dirac equation has spin
symmetry which means that $\Gamma(r)=A_{1}$ ($d\Gamma(r)/dr=0$) is
a constant, Eq. (7) has the following form
\begin{eqnarray}
&&\left\{\frac{d^2}{dr^2}-\frac{\kappa(\kappa+1)}{r^2}+\left[E+M-A_{1}\right]
\left[E-M-\Lambda(r)\right]\right\}F(r)=0\,,
\end{eqnarray}
and if the Dirac equation has pseudospin symmetry which means that
$\Lambda(r)=A_{2}$ ($d\Lambda(r)/dr=0$) is a constant, Eq. (8)
becomes
\begin{eqnarray}
&&\left\{\frac{d^2}{dr^2}-\frac{\kappa(\kappa-1)}{r^2}+\left[E-M-A_{2}\right]
\left[E+M-\Gamma(r)\right]\right\}G(r)=0\,.
\end{eqnarray}

\section{Nikiforov Uvarov Method}

The Nikiforov-Uvarov method could be used to solve a second-order
differential equation of the hypergeometric-type which can be
transformed by using appropriate coordinate transformation into
the following form
\begin{eqnarray}
\sigma^2(z)\frac{d^2\Psi(z)}{dz^2}+\sigma(z)\tilde{\tau}(z)\frac{d\Psi(z)}{dz}+\tilde{\sigma}(z)
\Psi(z)=0\,,
\end{eqnarray}
where $\sigma(z)$\,, and $\tilde{\sigma}(z)$ are polynomials, at
most, second degree, and $\tilde{\tau}(z)$ is a first-degree
polynomial. By taking the solution as
\begin{eqnarray}
\Psi(z)=\psi(z)\varphi(z)\,,
\end{eqnarray}
gives Eq. (11) as a hypergeometric type equation \cite{ref32}
\begin{eqnarray}
\frac{d^2\varphi(z)}{dz^2}+\frac{\tau(z)}{\sigma(z)}\frac{d\varphi(z)}{dz}+\frac{\lambda}{\sigma(z)}\,
\varphi(z)=0\,,
\end{eqnarray}
where $\psi(z)$ is defined by using the equation \cite{ref32}
\begin{eqnarray}
\frac{1}{\psi(z)}\frac{d\psi(z)}{dz}=\frac{\pi(z)}{\sigma(z)}\,,
\end{eqnarray}
and the other part of the solution in Eq. (12) is given by
\begin{eqnarray}
\varphi_{n}(z)=\frac{a_{n}}{\rho(z)}\frac{d^n}{dz^n}[\sigma^{n}(z)\rho(z)]\,,
\end{eqnarray}
where $a_{n}$ is a normalization constant, and $\rho(z)$ is the
weight function, and satisfies the following equation \cite{ref32}
\begin{eqnarray}
\frac{d\sigma(z)}{dz}+\frac{\sigma(z)}{\rho(z)}\frac{d\rho(z)}{dz}=\tau(z)\,.
\end{eqnarray}

The function $\pi(z)$ and the parameter $\lambda$ in the above
equation are defined as
\begin{eqnarray}
\pi(z)&=&\,\frac{1}{2}\,[\frac{d}{dz}\,\sigma(z)-\tilde{\tau}(z)]
\pm\bigg\{\frac{1}{4}\left[\frac{d}{dz}\,\sigma(z)-\tilde{\tau}(z)\right]^2
-\tilde{\sigma}(z)+k\sigma(z)\bigg\}^{1/2}\,,\\
\lambda&=&k+\frac{d}{dz}\,\pi(z)\,.
\end{eqnarray}

In the NU method, the square root in Eq. (17) must be the square
of a polynomial, so the parameter $k$ can be determined. Thus, a
new eigenvalue equation becomes
\begin{eqnarray}
\lambda=\lambda_{n}=-n\frac{d}{dz}\,\tau(z)-\frac{1}{2}\,(n^2-n)\frac{d^2}{dz^2}\,\sigma(z)\,.
\end{eqnarray}
and the derivative of the function
$\tau(z)=\tilde{\tau}(z)+2\pi(z)$ should be negative.

\section{Bound State Solutions}
\subsection{Hellmann Potential}
The Hellmann potential having the form
\begin{eqnarray}
V(r)=-\frac{a}{r}+\frac{b}{r}\,e^{-\beta r}\,,
\end{eqnarray}
has been used to explain the electron-ion \cite{ref33} or
electron-core interaction \cite{ref34}, alkali hydride molecules
and to study of inner-shell ionisation problem \cite{ref35}. We
present the plot of the above potential in Fig. (1) to see the
variation with position coordinate.

\textit{1. Spin Symmetric Solutions}

Inserting Eq. (20) into Eq. (9) and using the approximation given
in Eq. (1) instead of the spin-orbit coupling term, we obtain
\begin{eqnarray}
\left\{\frac{d^2}{dr^2}-\frac{\beta^2
\kappa(\kappa+1)}{(1-e^{-\beta r})^2}+\frac{\beta}{1-e^{-\beta
r}}\left(a-be^{-\beta r}\right)+\epsilon_{H}^{SS}\right\}F(r)=0\,,
\end{eqnarray}
where $H$ stands for the Hellmann potential and
$\epsilon_{H}^{SS}=(E+M-A_1)(E-M)$. Defining a new variable
$z=e^{-\beta r}$ and using the following abbreviations
\begin{subequations}
\begin{align}
&a^2_{1}=\kappa(\kappa+1)-\frac{1}{\beta^2}\left(a\beta+\epsilon_{H}^{SS}\right)\,,\\
&a^2_{2}=\frac{1}{\beta^2}\left[\beta(a+b)+2\epsilon_{H}^{SS}\right]\,,\\
&a^2_{3}=-\frac{1}{\beta^2}\left[b\beta+\epsilon_{H}^{SS}\right]\,,
\end{align}
\end{subequations}
we write Eq. (21) as
\begin{eqnarray}
\frac{d^2F(z)}{dz^2}+\frac{1-z}{z(1-z)}\frac{dF(z)}{dz}+\frac{1}{z^2(1-z)^2}\left[-a^2_{1}-a^2_{2}z-a^2_{3}z^2\right]F(z)=0\,,
\end{eqnarray}
Comparing the last equation with Eq. (11), we have
\begin{eqnarray}
\tilde{\tau}(z)=1-z\,,\,\,\,\,\,\sigma(z)=z(1-z)\,,\,\,\,\,\,
\tilde{\sigma}(z)=-a_1^2z^2-a_2^2z-a_3^2\,,
\end{eqnarray}
The function $\pi(z)$ is obtained from Eq. (17) as
\begin{eqnarray}
\pi(z)=\,-\frac{1}{2}\,z\,\mp\sqrt{(\,\frac{1}{4}\,+a_{3}^2-k)z^2+(a_{2}^2+k)z+a_{1}^2\,}\,,
\end{eqnarray}
The constant $k$ is determined by imposing a condition such that
the discriminant under the square root should be zero. The roots
of $k$ are $k_{1,2}=-a_{2}^2-2a_{1}^2\mp a_{1}(1+2\kappa)$.
Substituting the value of $k_{1}=-a_{2}^2-2a_{1}^2+
a_{1}(1+2\kappa)$ into Eq. (25), we get for $\pi(z)$
\begin{eqnarray}
\pi(z)(k \rightarrow k_{1})=\left\{
\begin{array}{lr}
-(a_{1}-\kappa)z+a_{1} & \\
-(1+\kappa-a_{1})z-a_{1}\,, & \\
\end{array}\right.
\end{eqnarray}
Now we calculate the polynomial $\tau(z)$  from $\pi(z)$ such that
its derivative with respect to $z$ must be negative. Thus we
obtain $\tau(z)$ for the second choice in last equation as
\begin{eqnarray}
\tau(z)=(2a_{1}-1)z-(1+2\kappa+2a_{1})\,,
\end{eqnarray}
The constant $\lambda$ in Eq. (18) becomes
\begin{eqnarray}
\lambda=-a^2_{2}-2a_{1}^2+a_{1}(1+2\kappa)-a_{1}+\kappa\,,
\end{eqnarray}
and Eq. (19) gives us
\begin{eqnarray}
\lambda_{n}=n(n-2a_{1})\,.
\end{eqnarray}
Substituting the values of the parameters given by Eq. (22), and
setting $\lambda=\lambda_n$, one can find the energy eigenvalues
for the Hellmann potential as
\begin{eqnarray}
E=\frac{1}{2}\left[A_1\mp
\sqrt{A_{1}^2-4(MA_{1}-M^2-N)\,}\right]\,,
\end{eqnarray}
where $N$ is a parameter written in terms of the quantum numbers
$n$ and $\kappa$ as
\begin{eqnarray}
N=-\frac{\beta^2}{4(n+\kappa)^2}\left[\frac{1}{\beta}(a-b)-(n^2-\kappa^2)-\kappa(\kappa+1)\right]^2-a\beta+\beta^2\kappa(\kappa+1)\,.
\end{eqnarray}
Now we find the upper component of the Dirac wave function. We
first compute the weight function from Eq. (16) with the help of
Eq. (27)
\begin{eqnarray}
\rho(z)=z^{-2(1+\kappa+a_{1})}\,(1-z)^{(1+2\kappa)}\,,
\end{eqnarray}
and we obtain from Eq. (15)
\begin{eqnarray}
\varphi_n(z)\sim
z^{-2(1+\kappa+a_{1})}\,(1-z)^{(1+2\kappa)}\,\frac{d^n}{dz^n}\,\left[
\,z^{n-\kappa-a_{1}-2}\,(1-z)^{n-2\kappa-1}\right]\,,
\end{eqnarray}
The polynomial solutions can be written in terms of the Jacobi
polynomials \cite{ref38}
\begin{eqnarray}
\varphi_n(z)\sim P_n^{(-2(1+\kappa+a_{1}),\,
-(1+2\kappa)\,)}\,(1-2z)\,.
\end{eqnarray}
The other part of the wave function is obtained from the Eq. (15)
as
\begin{eqnarray}
\psi(z)=z^{-a_{1}}(1-z)^{1-\kappa}\,,
\end{eqnarray}
Thus we write the upper component for the Hellmann potential in
Eq. (4) as
\begin{eqnarray}
F(z) \sim z^{-a_{1}}(1-z)^{1-\kappa} P_n^{(-2(1+\kappa+a_{1}),\,
-(1+2\kappa)\,)}\,(1-2z)\,,
\end{eqnarray}
By using Eq. (6a) and the identity for derivative of the Jacobi
polynomials given as
$\frac{d}{dx}P_{n}^{(p,q)(x)}=\frac{1}{2}(n+p+q+1)P_{n-1}^{(p+1,q+1)}(x)$
\cite{ref38}, we obtain the other component for the Hellmann
potential as
\begin{eqnarray}
G(z) &\sim
\frac{z^{-a_{1}}(1-z)^{1-\kappa}}{E+M-A}[\beta(\frac{1}{a_{1}}-\frac{\kappa}{ln
z })P_n^{(-2(1+\kappa+a_{1}),\,
-(1+2\kappa)\,)}\,(1-2z)\nonumber\\&-\frac{1}{4}(n-2a_{1})P_n^{(-(1+2\kappa+2a_{1}),\,
-(2+2\kappa)\,)}\,(1-2z)]\,.
\end{eqnarray}

\textit{2. Pseudospin Symmetric Solutions}

Inserting Eq. (20) into Eq. (10) and using the approximation given
in Eq. (1), we obtain
\begin{eqnarray}
\left\{\frac{d^2}{dr^2}-\frac{\beta^2
\kappa(\kappa-1)}{(1-e^{-\beta r})^2}+\frac{\beta}{1-e^{-\beta
r}}\left(a-be^{-\beta
r}\right)+\epsilon_{H}^{PSS}\right\}F(r)=0\,,
\end{eqnarray}
where $\epsilon_{H}^{PSS}=(E-M-A_2)(E+M)$. Using the same variable
and the following abbreviations
\begin{subequations}
\begin{align}
&a^2_{1}=\kappa(\kappa-1)-\frac{1}{\beta^2}\left(a\beta+\epsilon_{H}^{PSS}\right)\,,\\
&a^2_{2}=\frac{1}{\beta^2}\left[\beta(a+b)+2\epsilon_{H}^{PSS}\right]\,,\\
&a^2_{3}=-\frac{1}{\beta^2}\left[b\beta+\epsilon_{H}^{PSS}\right]\,,
\end{align}
\end{subequations}
we obtain
\begin{eqnarray}
\frac{d^2G(z)}{dz^2}+\frac{1-z}{z(1-z)}\frac{dG(z)}{dz}+\frac{1}{z^2(1-z)^2}\left[-a^2_{1}-a^2_{2}z-a^2_{3}z^2\right]G(z)=0\,,
\end{eqnarray}
Following the same steps in previous section, we write the energy
eigenvalues for the Hellmann potential for the case of pseudospin
symmetry
\begin{eqnarray}
E=\frac{1}{2}\left[A_2\mp
\sqrt{A_{2}^2+4(MA_{2}+M^2+N)\,}\right]\,,
\end{eqnarray}
where $N$ is given as
\begin{eqnarray}
N=-\frac{\beta^2}{4(n+\kappa)^2}\left[-\frac{1}{\beta}(a-b)+n^2+\kappa^2+\kappa(\kappa-3)\right]^2-a\beta+\beta^2\kappa(\kappa-1)\,.
\end{eqnarray}
and the lower component is written as
\begin{eqnarray}
G(z) \sim z^{-a_{1}}(1-z)^{1-\kappa} P_n^{(-2(1+\kappa+a_{1}),\,
-(1+2\kappa)\,)}\,(1-2z)\,.
\end{eqnarray}
Using Eq. (6b) gives us the other component as
\begin{eqnarray}
F(z) &\sim
\frac{z^{-a_{1}}(1-z)^{1-\kappa}}{M-E+A}[\beta(\frac{1}{a_{1}}-\frac{\kappa}{ln
z })P_n^{(-2(1+\kappa+a_{1}),\,
-(1+2\kappa)\,)}\,(1-2z)\nonumber\\&-\frac{1}{4}(n-2a_{1})P_n^{(-(1+2\kappa+2a_{1}),\,
-(2+2\kappa)\,)}\,(1-2z)]\,.
\end{eqnarray}
\subsection{Wei Hua Potential}
The Wei Hua potential is written
\begin{eqnarray}
V(r)=D\left[\frac{1-e^{-\beta r}}{1-ae^{-\beta r}}\right]^2\,,
\end{eqnarray}
which is proposed for bond-stretching vibration of diatomic
molecules \cite{ref36}. We give the plot of the Wei Hua potential
in Fig. (2).

\textit{1. Spin Symmetric Solutions}

Inserting last equation and Eq. (1) into Eq. (9), we obtain
\begin{eqnarray}
\left\{\frac{d^2}{dr^2}-\frac{\beta^2
\kappa(\kappa+1)}{(1-e^{-\beta r})^2}-D\left[\frac{1-e^{-\beta
r}}{1-ae^{-\beta r}}\right]^2+\epsilon_{WH}^{SS}\right\}F(r)=0\,,
\end{eqnarray}
where $WH$ stands for the Wei Hua potential and
$\epsilon_{WH}^{SS}=(E+M-A_{1})(E-M)$. Defining a new variable
$z=ae^{-\beta r}$, using the abbreviations
\begin{subequations}
\begin{align}
&a^2_{1}=\kappa(\kappa+1)-\frac{1}{\beta^2}\left(\epsilon_{WH}^{SS}-D\right)\,,\\
&a^2_{2}=-\frac{1}{\beta^2}\left[\frac{2D}{a}-2\epsilon_{WH}^{SS}\right]\,,\\
&a^2_{3}=-\frac{1}{\beta^2}\left[\epsilon_{WH}^{SS}-\frac{D}{a^2}\right]\,,
\end{align}
\end{subequations}
and following the same procedure in the above section for the
Hellmann potential, we write the energy eigenvalues of the Wei Hua
potential for the case of spin symmetry
\begin{eqnarray}
E=\frac{1}{2}\left[A_{1}\mp
\sqrt{A_{1}^2-4(MA_{1}-M^2-N)\,}\right]\,,
\end{eqnarray}
where $N$ is a parameter written in terms of the quantum numbers
$n$ and $\kappa$ as
\begin{eqnarray}
N=-\frac{\beta^2}{4(n+\kappa)^2}\left[n^2+\kappa^2+\kappa(\kappa+1)-\frac{2D}{\beta^2}(\frac{1}{a}-1)\right]^2+D+\beta^2\kappa(\kappa+1)\,.
\end{eqnarray}
and the lower component for the Wei Hua potential
\begin{eqnarray}
F(z) \sim z^{-a_{1}}(1-z)^{1-\kappa} P_n^{(-2(1+\kappa+a_{1}),\,
-(1+2\kappa)\,)}\,(1-2z)\,.
\end{eqnarray}
By using Eq. (6a) we obtain the other component as
\begin{eqnarray}
G(z) &\sim
\frac{z^{-a_{1}}(1-z)^{1-\kappa}}{E+M-A}[\beta(\frac{1}{a_{1}}-\frac{\kappa}{ln
z })P_n^{(-2(1+\kappa+a_{1}),\,
-(1+2\kappa)\,)}\,(1-2z)\nonumber\\&-\frac{1}{4}(n-2a_{1})P_n^{(-(1+2\kappa+2a_{1}),\,
-(2+2\kappa)\,)}\,(1-2z)]\,.
\end{eqnarray}
\textit{2. Pseudospin Symmetric Solutions}

Inserting Eqs. (45) and (1) into Eq. (10), we obtain
\begin{eqnarray}
\left\{\frac{d^2}{dr^2}-\frac{\beta^2
\kappa(\kappa-1)}{(1-e^{-\beta r})^2}-D\left[\frac{1-e^{-\beta
r}}{1-ae^{-\beta r}}\right]^2+\epsilon_{WH}^{PSS}\right\}F(r)=0\,,
\end{eqnarray}
$\epsilon_{WH}^{PSS}=(E-M-A_2)(E+M)$. Using the same variable $z$
for the Hellmann potential and defining the abbreviations
\begin{subequations}
\begin{align}
&a^2_{1}=\kappa(\kappa-1)-\frac{1}{\beta^2}\left(\epsilon_{WH}^{PSS}-D\right)\,,\\
&a^2_{2}=-\frac{1}{\beta^2}\left[\frac{2D}{a}-2\epsilon_{WH}^{PSS}\right]\,,\\
&a^2_{3}=-\frac{1}{\beta^2}\left[\epsilon_{WH}^{PSS}-\frac{D}{a^2}\right]\,,
\end{align}
\end{subequations}
and following the same procedure in the above section for the
Hellmann potential, we write the energy eigenvalues of the Wei Hua
potential for the case of pseudospin symmetry
\begin{eqnarray}
E=\frac{1}{2}\left[A_{2}\mp
\sqrt{A_{2}^2+4(MA_{2}+M^2+N)\,}\right]\,,
\end{eqnarray}
where $N$ is a parameter written in terms of the quantum numbers
$n$ and $\kappa$ as
\begin{eqnarray}
N=-\frac{\beta^2}{4(n+\kappa)^2}\left[n^2-\kappa^2+\kappa(\kappa-1)-\frac{2D}{\beta^2}(\frac{1}{a}-1)\right]^2+D+\beta^2\kappa(\kappa-1)\,.
\end{eqnarray}
and the upper component for the Wei Hua potential
\begin{eqnarray}
G(z) \sim z^{-a_{1}}(1-z)^{1-\kappa} P_n^{(-2(1+\kappa+a_{1}),\,
-(1+2\kappa)\,)}\,(1-2z)\,.
\end{eqnarray}
The other component can be obtained from Eq. (6b) as
\begin{eqnarray}
F(z) &\sim
\frac{z^{-a_{1}}(1-z)^{1-\kappa}}{M-E+A}[\beta(\frac{1}{a_{1}}-\frac{\kappa}{ln
z })P_n^{(-2(1+\kappa+a_{1}),\,
-(1+2\kappa)\,)}\,(1-2z)\nonumber\\&-\frac{1}{4}(n-2a_{1})P_n^{(-(1+2\kappa+2a_{1}),\,
-(2+2\kappa)\,)}\,(1-2z)]\,.
\end{eqnarray}
\subsection{Varshni Potential}
Varshni, for the first time, proposed the following potential
function
\begin{eqnarray}
V(r)=a\left[1-\frac{b}{r}\,e^{-\beta r}\right]\,,
\end{eqnarray}
to study the diatomic molecules \cite{ref37}. It is clearly seen
that the potential is very similar to the Hellmann potential which
could be seen in Fig. (3). All figures show that the form of the
potentials presented in this work are very similar.

Now we tend to study the spin and pseudospin symmetric solutions
of the Dirac equation for the above potential.

\textit{1. Spin Symmetric Solutions}

Inserting Eqs. (58) and Eq. (1) into Eq. (9), we obtain
\begin{eqnarray}
\left\{\frac{d^2}{dr^2}-\frac{\beta}{1-e^{-\beta
r}}\left(\frac{\beta\kappa(\kappa+1)}{1-e^{-\beta r}}-abe^{-\beta
r }\right)+\epsilon_{V}^{SS}\right\}F(r)=0\,,
\end{eqnarray}
where $V$ stands for the Varshni potential and
$\epsilon_{V}^{SS}=(E+M-A_{1})(E-M)-a$. Defining a new variable
$z=e^{-\beta r}$, using the abbreviations
\begin{subequations}
\begin{align}
&a^2_{1}=\kappa(\kappa+1)-\frac{\epsilon_{V}^{SS}}{\beta^2}\,,\\
&a^2_{2}=-\frac{1}{\beta^2}\left[ab\beta-2\epsilon_{V}^{SS}\right]\,,\\
&a^2_{3}=-\frac{1}{\beta^2}\left[\epsilon_{V}^{SS}-ab\beta\right]\,,
\end{align}
\end{subequations}
and following the same procedure in the above sections, we write
the energy eigenvalues of the Varshni potential for the case of
spin symmetry
\begin{eqnarray}
E=\frac{1}{2}\left[A_{1}\mp
\sqrt{A_{1}^2-4(MA_{1}-M^2-N)\,}\right]\,,
\end{eqnarray}
where
\begin{eqnarray}
N=-\frac{\beta^2}{4(n+\kappa)^2}\left[-\frac{ab}{\beta}+n^2+\kappa^2+\kappa(\kappa+1)\right]^2+\beta^2\kappa(\kappa+1)+a\,.
\end{eqnarray}
and the lower component for the Varshni potential
\begin{eqnarray}
F(z) \sim z^{-a_{1}}(1-z)^{1-\kappa} P_n^{(-2(1+\kappa+a_{1}),\,
-(1+2\kappa)\,)}\,(1-2z)\,.
\end{eqnarray}
By using Eq. (6a) we obtain the other component as
\begin{eqnarray}
G(z) &\sim
\frac{z^{-a_{1}}(1-z)^{1-\kappa}}{E+M-A}[\beta(\frac{1}{a_{1}}-\frac{\kappa}{ln
z })P_n^{(-2(1+\kappa+a_{1}),\,
-(1+2\kappa)\,)}\,(1-2z)\nonumber\\&-\frac{1}{4}(n-2a_{1})P_n^{(-(1+2\kappa+2a_{1}),\,
-(2+2\kappa)\,)}\,(1-2z)]\,.
\end{eqnarray}
\textit{2. Pseudospin Symmetric Solutions}

Inserting Eqs. (58) and (1) into Eq. (10), we obtain
\begin{eqnarray}
\left\{\frac{d^2}{dr^2}-\frac{\beta}{1-e^{-\beta
r}}\left(\frac{\beta\kappa(\kappa-1)}{1-e^{-\beta r}}-abe^{-\beta
r }\right)+\epsilon_{V}^{PSS}\right\}F(r)=0\,,
\end{eqnarray}
$\epsilon_{V}^{PSS}=(E-M-A_{2})(E+M)$. Using the same variable $z$
for the Hellmann potential and defining the abbreviations
\begin{subequations}
\begin{align}
&a^2_{1}=\kappa(\kappa-1)-\frac{\epsilon_{V}^{PSS}}{\beta^2}\,,\\
&a^2_{2}=-\frac{1}{\beta^2}\left[ab\beta-2\epsilon_{V}^{PSS}\right]\,,\\
&a^2_{3}=-\frac{1}{\beta^2}\left[\epsilon_{V}^{PSS}-ab\beta\right]\,,
\end{align}
\end{subequations}
and following the same procedure in the above sections, we write
the energy eigenvalues of the Varshni potential for the case of
pseudospin symmetry
\begin{eqnarray}
E=\frac{1}{2}\left[A_{2}\mp
\sqrt{A_{2}^2+4(MA_{2}+M^2+N)\,}\right]\,,
\end{eqnarray}
where
\begin{eqnarray}
N=-\frac{\beta^2}{4(n+\kappa)^2}\left[n^2+\kappa^2+\kappa(\kappa-3)-\frac{ab}{\beta}\right]^2-\beta^2\kappa(\kappa-1)+a\,.
\end{eqnarray}
and the upper component for the Varshni potential
\begin{eqnarray}
G(z) \sim z^{-a_{1}}(1-z)^{1-\kappa} P_n^{(-2(1+\kappa+a_{1}),\,
-(1+2\kappa)\,)}\,(1-2z)\,.
\end{eqnarray}
Using Eq. (6b) gives the other component as
\begin{eqnarray}
F(z) &\sim
\frac{z^{-a_{1}}(1-z)^{1-\kappa}}{M-E+A}[\beta(\frac{1}{a_{1}}-\frac{\kappa}{ln
z })P_n^{(-2(1+\kappa+a_{1}),\,
-(1+2\kappa)\,)}\,(1-2z)\nonumber\\&-\frac{1}{4}(n-2a_{1})P_n^{(-(1+2\kappa+2a_{1}),\,
-(2+2\kappa)\,)}\,(1-2z)]\,.
\end{eqnarray}

\section{Results and Discussions}
We have listed some numerical values for energy eigenvalues in
Tables I-VI for the cases of spin and pseudospin symmetries,
separately. We have used the same parameter values in both of spin
and pseusospin symmetric cases for the Hellmann potential, this is
valid also for the Varshni potential. But the values of the
parameters for Wei Hua potential are different for the cases of
spin and pseudospin symmetries. It could be seen that the
dependence of the bound states for the Wei Hua potential are more
sensitive. It also should be stressed that the spin (and
pseudospin) doublets, i.e., $(0,-2)$ and $(0,1)$ states or
$(1,-2)$ and $(1,1)$ states, etc. could be seen up to fourth
decimal in energy eigenvalues.

\section{Conclusion}
We have studied the approximate bound state solutions of the Dirac
equation for the Hellmann potential, Wei Hua potential and Varshni
potential, which have an exponential form depending on the
spatially coordinate $r$, for the cases where the Dirac equation
has pseudospin and spin symmetry, respectively. The variation of
the above potentials according to coordinate $r$ are given in Figs
I-III. We have obtained the energy eigenvalue equations and the
related two-component spinor wave functions with the help of
Nikiforov-Uvarov method and summarized the numerical results for
the bound states in Tables I-VI. It is also seen that the
Nikiforov-Uvarov method is a suitable method to study the bound
state solutions of the above potentials.

\section{Acknowledgments}
This research was partially supported by the Scientific and
Technical Research Council of Turkey.


\newpage

\begin{table}[htp]
\centering \caption{The energy eigenvalues of the Hellmann
potential for the case of spin symmetry for $a=0.25, b=0.20,
\beta=0.02, A_{1}=M=10$.}
\begin{tabular}{@{}ccccccc@{}}
$\ell$ & $n$ & $\kappa$ & $E>0$ & $n$ & $\kappa$ & $E>0$ \\ \hline
1 & 0 & -2 & 9.9995294 & 0 & 1 & 9.9995575\\
2 &   & -3 & 9.9997604 & & 2 & 9.9997394\\
3 &   & -4 & 9.9999536 & & 3 & 9.9999700\\
4 &   & -5 & 10.0002770 & & 4 & 10.0002900\\
1 & 1 & -2 & 9.9994575 & 1 & 1 & 9.9995700\\
2 &   & -3 & 9.9996894 & & 2 & 9.9997300\\
3 &   & -4 & 9.9999464 & & 3 & 9.9999700\\
4 &   & -5 & 10.0002740 & & 4 & 10.0002900\\
\end{tabular}
\end{table}

\begin{table}[htp]
\centering \caption{ The energy eigenvalues of the Hellmann
potential for the case of pseudospin symmetry for $a=0.25, b=0.20,
\beta=0.02, A_{2}=M=10$.}
\begin{tabular}{ccccccc}
$\ell$ & $n$ & $\kappa$ & $E>0$ & $n$ & $\kappa$ & $E>0$ \\ \hline
1 & 0 & -2 & 9.9998031 & 0 & 1 & 9.9997925\\
2 &   & -3 & 9.9997710 & & 2 & 9.9998598\\
3 &   & -4 & 9.9997412 & & 3 & 9.9998977\\
4 &   & -5 & 9.9997125 & & 4 & 9.9992950\\
1 & 1 & -2 & 9.9993925 & 1 & 1 & 9.9998281\\
2 &   & -3 & 9.9994514 & & 2 & 9.9998599\\
3 &   & -4 & 9.9994310 & & 3 & 9.9999016\\
4 &   & -5 & 9.9993933 & & 4 & 9.9999477\\
\end{tabular}
\end{table}

\begin{table}[htp]
\centering \caption{ The energy eigenvalues of the Wei Hua
potential for the case of spin symmetry for $a=0.10, D=0.0001,
\beta=0.01, A_{1}=2, M=0.001$.}
\begin{tabular}{ccccccc}
$\ell$ & $n$ & $\kappa$ & $E>0$ & $n$ & $\kappa$ & $E>0$ \\ \hline
1 & 0 & -2 & 1.9986997 & 0 & 1 & 1.9963313\\
2 &   & -3 & 1.9993378 & & 2 & 1.9991501\\
3 &   & -4 & 1.9995723 & & 3 & 1.9996379\\
4 &   & -5 & 1.9996860 & & 4 & 1.9997974\\
1 & 1 & -2 & 1.9976352 & 1 & 1 & 1.9985369\\
2 &   & -3 & 1.9993378 & & 2 & 1.9996379\\
3 &   & -4 & 1.9994823 & & 3 & 1.9996379\\
4 &   & -5 & 1.9994378 & & 4 & 1.9998700\\
\end{tabular}
\end{table}

\begin{table}[htp]
\centering \caption{ The energy eigenvalues of the Wei Hua
potential for the case of pseudospin symmetry for $a=0.25, D=0.01,
\beta=0.10, A_{2}=10, M=1$.}
\begin{tabular}{ccccccc}
$\ell$ & $n$ & $\kappa$ & $E>0$ & $n$ & $\kappa$ & $E>0$ \\ \hline
1 & 0 & -2 & 1.0049979 & 0 & 1 & 0.9956234\\
2 &   & -3 & 1.0056224 & & 2 & 1.0024995\\
3 &   & -4 & 1.0057785 & & 3 & 1.0039570\\
4 &   & -5 & 1.0058222 & & 4 & 1.0045295\\
1 & 1 & -2 & 1.0006250 & 1 & 1 & 1.0000000\\
2 &   & -3 & 0.9974994 & & 2 & 1.0024763\\
3 &   & -4 & 0.9952527 & & 3 & 1.0045295\\
4 &   & -5 & 0.9932780 & & 4 & 1.0064216\\
\end{tabular}
\end{table}

\begin{table}[htp]
\centering \caption{ The energy eigenvalues of the Varshni
potential for the case of spin symmetry for $a=b=0.15,
\beta=0.001, A_{1}=M=5$.}
\begin{tabular}{ccccccc}
$\ell$ & $n$ & $\kappa$ & $E>0$ & $n$ & $\kappa$ & $E>0$ \\ \hline
1 & 0 & -2 & 4.9999970 & 0 & 1 & 4.9999814\\
2 &   & -3 & 5.0000009 & & 2 & 4.9999992\\
3 &   & -4 & 5.0000023 & & 3 & 5.0000024\\
4 &   & -5 & 5.0000030 & & 4 & 5.0000034\\
1 & 1 & -2 & 4.9999884 & 1 & 1 & 4.9999961\\
2 &   & -3 & 5.0000070 & & 2 & 5.0000050\\
3 &   & -4 & 5.0000022 & & 3 & 5.0000024\\
4 &   & -5 & 5.0000023 & & 4 & 5.0000036\\
\end{tabular}
\end{table}

\begin{table}[htp]
\centering \caption{ The energy eigenvalues of the Varshni
potential for the case of pseudospin symmetry for $a=b=0.15,
\beta=0.001, A_{2}=M=5$.}
\begin{tabular}{ccccccc}
$\ell$ & $n$ & $\kappa$ & $E>0$ & $n$ & $\kappa$ & $E>0$ \\ \hline
1 & 0 & -2 & 5.0000001 & 0 & 1 & 4.9999908\\
2 &   & -3 & 5.0000008 & & 2 & 4.9999984\\
3 &   & -4 & 5.0000009 & & 3 & 5.0000001\\
4 &   & -5 & 5.0000008 & & 4 & 5.0000008\\
1 & 1 & -2 & 4.9999995 & 1 & 1 & 4.9999979\\
2 &   & -3 & 5.0000007 & & 2 & 4.9999994\\
3 &   & -4 & 5.0000004 & & 3 & 5.0000002\\
4 &   & -5 & 5.0000000 & & 4 & 5.0000008\\
\end{tabular}
\end{table}

\newpage

\begin{figure}
\begin{center}
\includegraphics[height=3in, width=5in, angle=0]{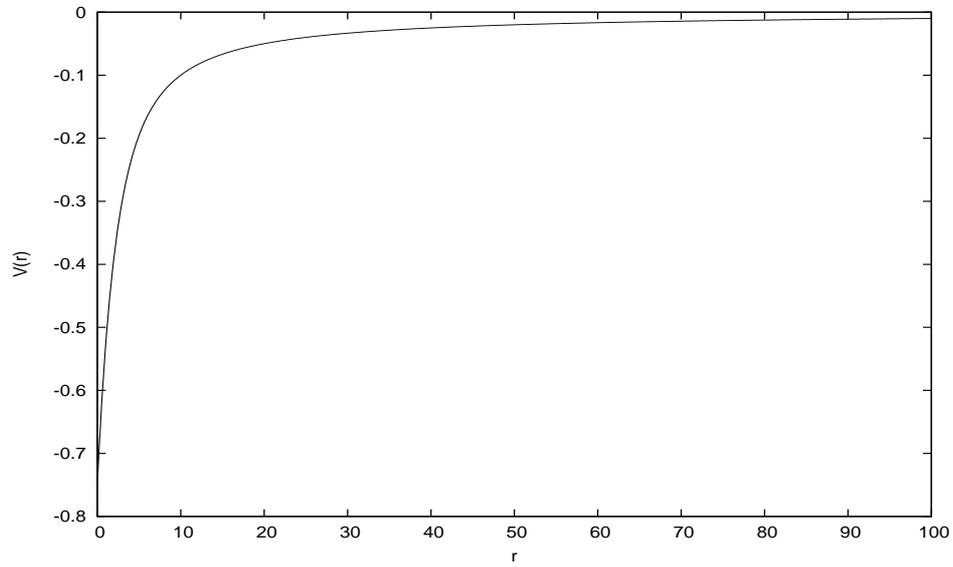}
\caption{The Hellman potential.}
\end{center}
\end{figure}

\begin{figure}
\begin{center}
\includegraphics[height=3in, width=5in, angle=0]{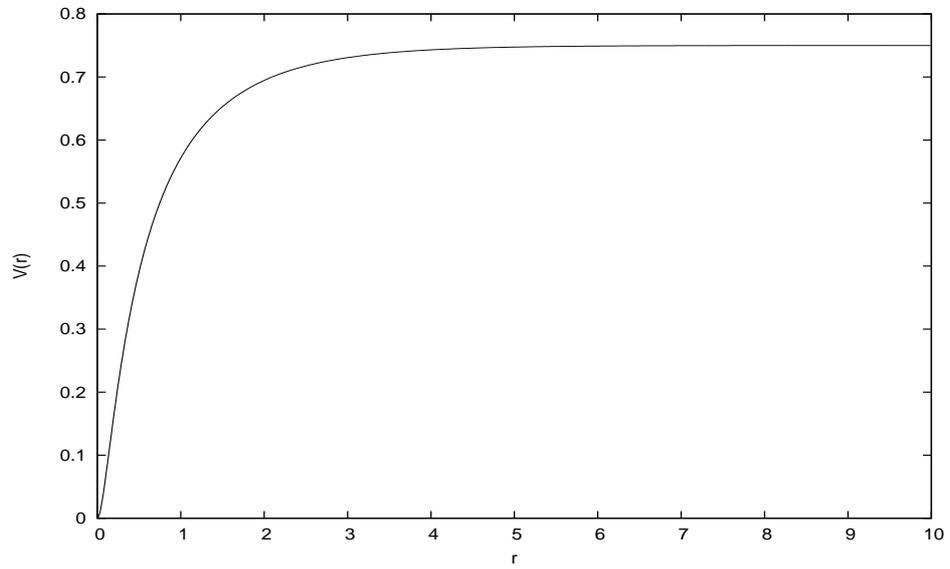}
\caption{The Wei Hua potential.}
\end{center}
\end{figure}

\begin{figure}
\begin{center}
\includegraphics[height=3in, width=5in, angle=0]{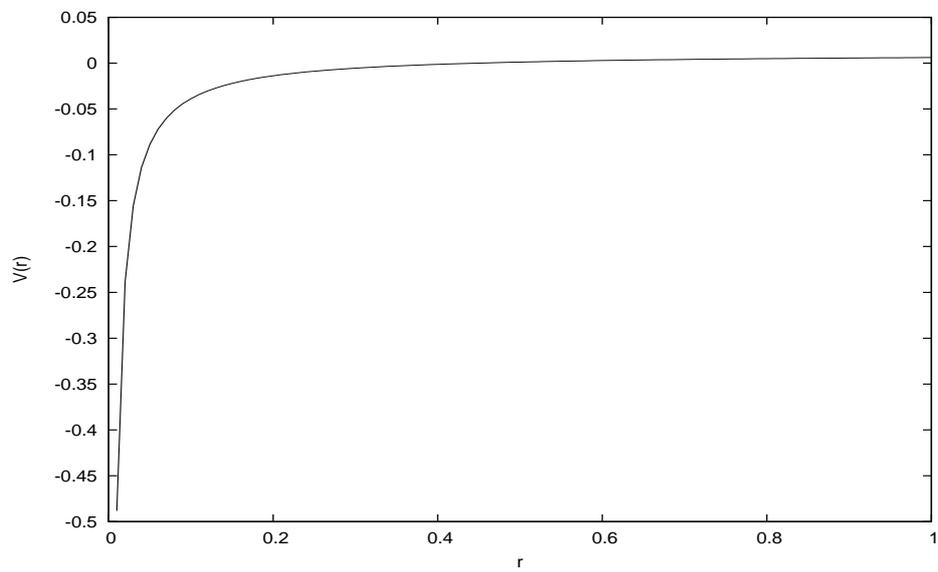}
\caption{The Varshni potential.}
\end{center}
\end{figure}

\end{document}